\documentclass[smallextended,final]{svjour3}

\usepackage{amsmath} \usepackage{graphicx} \usepackage{cite}

\begin{document}
\renewcommand{\makeheadbox}{}

\title{Towards a systematic characterization of the potential of
  demand side management}%
\titlerunning{Systematic characterization of the potential of demand
  side management}

\author{David Kleinhans}
\institute{D.\ Kleinhans
  \at NEXT ENERGY -- EWE Research Centre for Energy Technology at the
  University of Oldenburg, Carl-von-Ossietzky-Stra{\ss}e 15, 26129
  Oldenburg, Germany\\
  \email{david.kleinhans@next-energy.de}}

 \date{}
\maketitle

\begin{abstract}
  With an increasing share of electric energy produced from
  non-dis\-patch\-able renewable sources both energy storage and
  demand side management might gain tremendously in importance. While
  there has been significant progress in general properties and
  technologies of energy storage, the systematic characterization of
  features particular to demand side management such as its
  intermittent, time-dependent potential seems to be lagging
  behind. As a consequence, the development of efficient and
  sustainable strategies for demand side management and its
  integration into large-scale energy system models are impeded.

  This work introduces a novel framework for a systematic
  time-resolved characterization of the potential for demand side
  management. It is based on the specification of individual devices
  both with respect to their scheduled demand and their potential of
  load shifting.  On larger scales sector-specific profiles can
  straightforwardly be taken into account. The potential for demand
  side management is then specified in terms of size and capacity of
  an equivalent storage device. This intermediate layer of abstraction
  isolates the gross effects and opportunities of demand side
  management from questions concerning the operation of individual
  devices, which eases the integration of the results in existing
  schemes for the commitment of production units and system models.

  The novel framework is applicable for conceptual or predictive
  approaches and for real-time optimization and operation of energy
  systems involving demand side management.

\end{abstract}

\section{\label{sec:introduction}Introduction}
Recent studies emphasize that the energy demand of future large scale
power systems can be supplied from renewable sources
\cite{Jacobson2009,Delucchi2011}. The increased integration of energy
converters based on renewable sources, however, exhibits a tremendous
challenge for the operation of power systems, since the production
from the most extensive resources solar and wind is intermittent and
therefore not flexibly dispatchable
\cite{Georgilakis2008,Rodriguez2013}. Consequently their contributions
to the balancing of the power systems and to ancillary services are
limited in comparison to today's power plants
\cite{Rasmussen2012,Genoese2013}. For a large-scale integration of
conversion technologies based on renewable sources as planned in
Europe until 2050 therefore alternative technologies for balancing and
ancillary services need to be considered, installed, and
integrated. From the current perspective the most promising options
are energy storages and demand side management
\cite{Stotzer2011,Heide2010,Strbac2008}, since these technologies
allow for a shifting of electricity provision or consumption in time
and, hence, can contribute to a significant reduction in required
balancing energy and power.

In the context of this contribution \emph{demand side management}
(DSM) is used as a proxy for technologies for the short term
coordinated control and operation of loads at consumers'
side.\footnote{Equivalent expressions e.g.\ would have been
  \emph{demand response} or \emph{load management}. Strategies for the
  more long term implementation of improvements in energy efficiency,
  which sometimes are also associated with \emph{demand side
    management} \cite{Loughran2004}, however are not considered.}
Prominent examples for the application of DSM in the domestic sector
are e.g.\ electric heating devices operated either via power line or
radio communication
\cite{Strbac2008,Pollhammer2011,Arteconi2012,Arteconi2012a}. Large-scale
consumers from the industrial sector often sign individual agreements
with the grid operators on their contribution to DSM.  While DSM
currently is mainly employed in cases of large unbalances in the grid
or in order to increase the overall efficiency of unit commitment
schemes \cite{Stadler2008a}, its role might change in future. For
future systems based first and foremost on renewable energy sources,
energy storage and DSM techniques are one option for the instantaneous
adjustment of the residual load \cite{Stadler2008a}.  As a component
of systemic importance which e.g.\ in so-called \emph{smart grids}
might even be controlled autonomously \cite{Palensky2011}, a need for
the quantitative characterization of the instantaneous potential for
DSM arises.  Although DSM already today is a substantial component of
procedures for the optimization of power systems
\cite{Mathieu2012,Gayme2011,Heussen2010,Stotzer2011,Pollhammer2011}
both with respect to system stability and operating costs,\footnote{A
  general formulation applicable to complex power systems, which also
  involves demand side management at the respective nodes, is e.g.\
  provided in \cite{Heussen2010}.} its large-scale potential has not
been characterized and investigated systematically.

In a number of recent contributions on energy system modeling general
aspects of the integration of renewable sources were addressed from a
conceptual perspective, with a focus on the gross interactions between
energy storages, the extension of the transmission grids and
strategies for the expansion of conversion capacities for renewable
sources (see e.g.\
\cite{Rodriguez2013,Heide2010,Heide2011,MurthyBalijepalli2011,Genoese2013}).
Technically storage solutions and DSM technologies often are not
differentiated in these approaches
\cite{DeJonghe2012,Milano2010,Rasmussen2012}. While there from a
modeling perspective indeed might be analogies between energy storage
and DSM \cite{Palensky2011}, fundamental differences are evident. In
contrast to traditional storage devices with DSM rather complicated
preconditions have to be considered, since the restrictions of the
individual loads with respect to their time frame of management and
need to be taken into account. Generally this results in time
dependent buffer sizes and capacities. How can these restriction be
simplified, formalized and efficiently be integrated into energy
system models?

The present work aims to develop a framework for the characterization
of the instantaneous potential for DSM at intermediate complexity
suitable for the integration in large scale simulations. Similarly to
the characterization of energy storage solutions the DSM potential is
determined by the properties of storage-equivalent energy buffers, but
now with explicit time dependence.  At this stage practical problems
involving e.g.\ the control of individual devices and rebound effects
by intention are avoided, since these questions need to be addressed
and solved independently from the general concept of DSM integration.
As a consequence the resulting potential of DSM much more
straightforwardly can be interpreted and integrated into conceptual
simulation of future power systems than the current state of the art,
which treats DSM mainly from a unit-commitment-perspective (see e.g.\
\cite{Stotzer2011,Pollhammer2011}).

For reasons of clarity, throughout the work the \emph{size} of energy
buffers refers to their volume in terms of energy, whereas their
\emph{capacity} characterizes the maximum power of production or
consumption.

\section{\label{sec:novel-fram-quant}Novel framework for the
  quantitative characterization of demand side management}
Generally two classes of DSM can be differentiated: (1) loads which
can be \emph{shifted} in time without a significant change in the
gross energy consumption and (2) loads which can be reduced without
replacement (\emph{curtailment}). The first class e.g.\ involves a
broad range of heating and cooling processes, where shifting the
cycles in time only has a negligible impact on consumers. Also
dispatchable devices such as e.g.\ domestic white goods belong to this
class.  DSM actions of the 2nd class such as e.g.\ requests for the
temporal reduction of production without replacement generally have a
huger impact on the electricity consumers' interests, since the
overall availability of energy is directly affected.  For the core
part of the present paper we will restrict to DSM class 1, where
management and distribution of the energy demand is the most relevant
issue. Aspects for the integration of class 2 DSM in the framework are
discussed briefly in the context of Appendix
\ref{sec:integration-class-2}. For reasons of clarity we henceforth
use the time $t$ as a continuous variable. For correspondent
expression applicable in case of discrete time steps as often used in
simulations we refer to Appendix \ref{sec:time-discr-form}. Constants
and variables used for parametrization are listed in Table
\ref{tab:nomenclature}.

\begin{table}
  \centering
  \begin{tabular}{llll}\hline
    Variable&Description&Unit&Constraints\\\hline\hline
    $C$&Number of categories&---&$\ge 1$\\
    $\Delta t_c$&Time frame of management&[h]&$\ge 0$\\
    $L_c(t)$&Scheduled load for category $c$&[MW]&$\ge 0$\\
    $\Lambda_c(t)$&Maximum load / capacity for category $c$ &[MW]&$\ge 0$\\
    $R_c(t)$&Realized load (after DSM) for category $c$&[MW]&$\ge 0$, $\le \Lambda_c(t)$\\
    $P_c[R_c](t)$&Charge rate DSM category $c$ for $R_c(t)$&[MW]&see Eqn. (\ref{eq:validity-r})\\
    $E_c[R_c](t)$&Energy content category $c$ for $R_c(t)$&[MWh]&see Eqn. (\ref{eq:validity-r})\\
    \hline
  \end{tabular}
  \caption{\label{tab:nomenclature}Parameters and variables used for
    characterization of the DSM potential in Section
    \ref{sec:novel-fram-quant} with their
    respective units and constraints.}
\end{table}

Consumers or loads eligible for DSM class 1 can be characterized with
respect to their potential for contributions to DSM. In order to ease
the scalability of the procedure a number of $C$ \emph{categories}
$c=1,...,C$ is introduced, each specified by means of the maximum
allowed time frame for management, $\Delta t_c$, in units of
hours. The respective consumer loads then need to be apportioned among
the categories in a sense, that time resolved profiles of scheduled
loads $L_c(t)$ and maximum loads $\Lambda_c(t)$ can be aggregated for
the each category, both in units of MW. Here $\Lambda_c(t)$ is defined
as the load capacity (i.e.\ the maximal realizable load) available for
DSM in category $c$ at time $t$.  The data can e.g.\ stem from
simulations, predictions, experience or standard load
profiles. Examples for the application to different load profiles and
capacities are provided and discussed in Section
\ref{sec:examples}. From the individual loads $L_c(t)$ the cumulative
scheduled load $L$ can be obtained as
\begin{equation}
  \label{eq:total-load}L(t)=\sum_{c=1}^CL_c(t)\quad.
\end{equation}

With DSM deviations from the scheduled loads can be enforced resulting
in new time series $R_c(t)$ of \emph{realized loads} in the respective
categories.\footnote{With \emph{realized loads} we here refer to the
  loads after the application of DSM. Discrepancies between scheduled
  loads and the actual demand are not taken into account at this
  stage.}  For interpretation of DSM as an energy storage-equivalent
action, the time-resolved energy balance of the DSM actions can be
investigated. If the scheduled loads are considered as a reference, we
can figuratively consider an energy buffer of category $c$ to be
\emph{charged} if $R_c(t)>L_c(t)$. On the contrary, the buffer is
\emph{discharged} if $R_c(t)<L_c(t)$.  The utilization of DSM in terms
of energy buffers in the respective categories can then be determined
by the charge rate
\begin{equation}
  \label{eq:charge-rate}
  P_c[R_c(t)](t)=R_c(t)-L_c(t)\quad.
\end{equation}
If we define the corresponding energy buffers to be empty at $t=0$,
their energy content (or filling levels) can be calculated as integral
of the charge rates,
\begin{equation}
  \label{eq:filling-level}
  E_c[R_c(t)](t)=  \int_{0}^{t}\mathrm{d}t'\ P_c[R_c(t)](t')\quad.
\end{equation}
The square brackets indicate, that the expressions depend explicitly
on the time series of realized loads in the respective categories.

In this terminology storage-equivalent buffers of DSM in the
respective categories can now be specified by means of time-dependent
envelopes of the sizes and the capacities. Functions for the upper
(max) and lower (min) envelopes can be defined from the respective
time series for the scheduled and maximal loads by means of
\begin{subequations}
  \label{eq:e-and-p-definitions}
  \begin{eqnarray}
    \label{eq:e-max-ts}E^{\max}_c(t)&:=&E_c[R_c(t+\Delta t_c)](t)=\int_{t}^{t+\Delta t_c}\mathrm{d}t'\ L_c(t')\quad,\\
    \label{eq:e-min-ts}E^{\min}_c(t)&:=&E_c[R_c(t-\Delta t_c)](t)=-\int_{t-\Delta t_c}^{t}\mathrm{d}t'\ L_c(t')\quad,\\
    \label{eq:p-plus}
    P^{\max}_c(t)&:=&\Lambda_c(t)-L_c(t)\quad, \mbox{and}\\
    \label{eq:p-minus}
    P^{\min}_c(t)&:=&-L_c(t)\quad.
  \end{eqnarray}
\end{subequations}
In (\ref{eq:e-max-ts}) it is assumed, that all loads in the respective
categories are preponed by their maximum time frame of management
corresponding to the maximum possible charge of the associated energy
buffer. The time-dependent maximum charge is, hence, determined by the
energy of upcoming loads within the time frame of management. The
opposite case, i.e.\ loads delayed to their latest possible moment of
realization, determines a time dependent lower limit for the buffer
size, $E^{\min}_c(t)$, as defined in Equation
(\ref{eq:e-min-ts}). Equations (\ref{eq:p-plus}) and
(\ref{eq:p-minus}) specify the upper and lower envelopes for charging
and discharging the respective buffers from time series of the
scheduled and the maximum acceptable loads.

The resulting degrees of freedom in operation are the representations
of the realized loads $R_c(t)$ in the respective categories. From the
considerations above the following requirements for the validity of
these functions can be formulated:
\begin{subequations}
  \label{eq:validity-r}
  \begin{eqnarray}
    \label{eq:validity-r-size}
    E_c^{\min}(t)\le &E_c[R_c(t)](t)&\le E^{\max}_c(t)\quad\forall t,c\quad\mbox{and}\\
    \label{eq:validity-r-capacity}
    P_c^{\min}(t)\le &P_c[R_c(t)](t)& \le P_c^{\max}(t)\quad\forall t,c\quad.
  \end{eqnarray}
\end{subequations}
Equation (\ref{eq:validity-r-size}) guarantees, that the charge state
of the storage-equivalent buffers in the respective categories does
not exceed its limits, whereas Equation (\ref{eq:validity-r-capacity})
accounts for the limited charging capacities. Apart from the explicit
dependence on time these parameters correspond to the restrictions of
storage devices, which emphasizes a formal analogy to energy storage
devices and eases the practical implementation of DSM in relevant
models and procedures.

Two examples for the application of this framework and the graphical
presentation and interpretation of the results are detailed in the
subsequent section.

\section{\label{sec:examples}Examples for application}
\subsection{\label{sec:schematic-example}Schematic example}
\begin{figure}
  \centering
  \includegraphics[width=.99\textwidth]{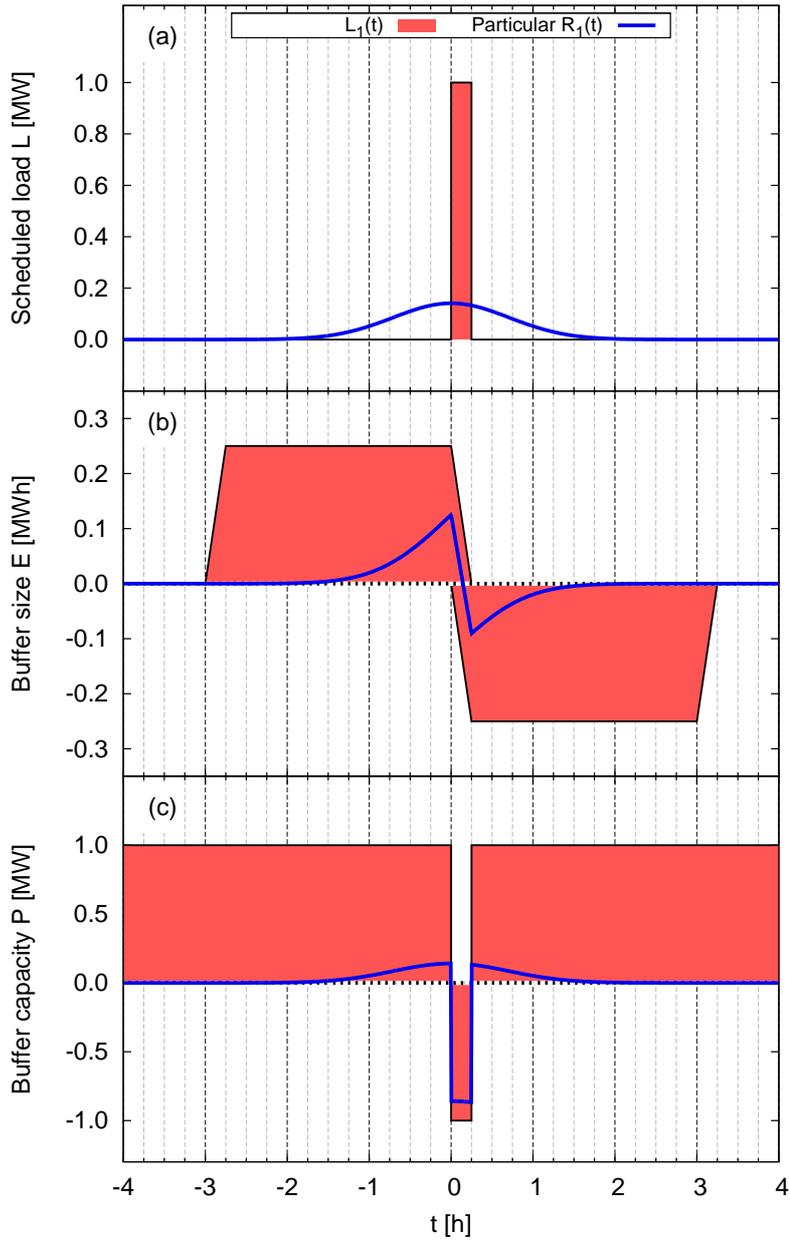}
  \caption{\label{fig:example-1-schematic}Application of the framework
    for characterization of the potential for demand side management
    to a single peak of a load eligible for demand side management
    with a time frame of $3~\mathrm{h}$. The maximum capacity just as
    the peak load was set to $1$. The panels indicate the scheduled
    load (a) and the envelopes of the size (b) and the capacity (c) of
    a storage-equivalent buffer as defined in Equation
    (\ref{eq:e-and-p-definitions}). The blue line exemplifies the
    charge state $E_1[R_1](t)$ (panel b) and the charge rate
    $P_1[R_1](t)$ (panel c) for a specific representation of the
    realized load $R_1(t)$ plotted in panel (a).}
\end{figure}
As a first, schematic example a system with only one category (i.e.\
$C=1$) and a single load pulse eligible for DSM is
considered. Specifically we use $\Delta t_1=3$, $\Lambda_1(t)=1$, and
\begin{equation}
  L_1(t)=\begin{cases}
    1&0<t<0.25\\
    0&\mbox{otherwise}
  \end{cases}\quad,
\end{equation}
reflecting a single load pulse of $15\ \mathrm{min}$ duration and
amplitude $1$ (see also panel (a) of Figure
\ref{fig:example-1-schematic}).  The scheduled load and the
corresponding envelopes of buffer sizes and capacities as calculated
from Equations (\ref{eq:e-and-p-definitions}) are exhibited in Figure
\ref{fig:example-1-schematic}.

From inspection of the figure the idea of the interpretation in terms
of a storage-equivalent energy buffer can become clear. The reference
is the scheduled load, $L_1(t)$. The envelope of the buffer size is
determined by the curves $E^{\max}_1(t)$ and $E^{\min}_1(t)$ as
defined in Equations (\ref{eq:e-and-p-definitions}). With the single
peak considered here the energy balance of any valid realization with
respect to the scheduled load by construction can be only positive at
$t<0$ and only negative at $t>0.25$, since the scheduled load is
non-zero only for $t\in[0,0.25]$ implying a rather limited flexibility
for DSM. This is reflected by the envelope in panel (b) of Figure
\ref{fig:example-1-schematic}. In the last panel the envelope of the
charge and discharge rates is plotted. As a striking feature the
buffer can only be discharged in the interval $0<t<0.25$, since the
scheduled load is $0$ otherwise.

Any realized load curve $R_1(t)$, that stays within the envelopes
defined by the extremal sizes and capacities is a valid. Formally this
criterion is provided by Equations (\ref{eq:validity-r}). In the three
panels of Figure \ref{fig:example-1-schematic} as an example the
resulting charge state and rate for
\begin{equation}
  R_1(t)=\left\{\begin{array}{ll}\mathcal{N}e^{-t^2}&|t|\le 3\\0&\mbox{otherwise}\end{array}\right.
\end{equation}
with the normalization constant
\begin{equation}
  \mathcal{N}:=0.25\left(\int
    _{-3}^{3}\mathrm{d}t\,e^{-t^2}\right)^{-1}\approx 0.141
\end{equation} are plotted. $R_1(t)$ is a valid representation of
the realized load, since it meets the constraints defined in Equation
(\ref{eq:validity-r}) as also shown in panels (b) and (c) of the
figure.  From panel (b) it becomes evident, that the
corresponding $E_1[R_1](t)$ fits in the rather odd envelope of the
buffer size. The charge rates have discontinuities at
$t=0$ and $t=0.25$, which is due to discontinuities in the scheduled
load $L_1(t)$.

\subsection{\label{sec:char-nati-potent}Characterization of the
  national potential of a European country for demand side management}
\begin{table}
  \centering
  \begin{minipage}{15cm}  
    \renewcommand\footnoterule{}
    \begin{tabular}{lllrr}\hline
      c&Code&Description&\multicolumn{1}{c}{$\Lambda$ [MW]}&\multicolumn{1}{c}{$\Delta t$ [h]}\\\hline\hline
      1&HEAT&Electric Heating&25000&8\\
      2&AC/W&AC and hot water&10000&4-8\footnote{For analysis $\Delta t_2=6\ \mathrm{h}$ is considered. }\\
      3&CO-D&Domestic cooling devices&2000&1\\
      4&WHIT&Domestic white goods&1400&24\\
      5&VENT&Ventilation&900&4\\
      6&CO-L&Cooling (low and intermediate power facilities)&800&4\\
      7&CO-H&Cooling (high power facilities)&200&4\\
      8&IN-1&Industry (cross-sectional technologies)&800&4\\
      9&IN-2&Industry (high demand, curtailment only)&2200&-\\\hline\\[-5.3ex]
    \end{tabular}
  \end{minipage}
  \caption{\label{tab:klobasa-categories}Categorization of loads and their corresponding
    potential for demand side management as compiled and estimated by Klobasa 
    \cite[p.\ 133]{Klobasa2007} (categories numbers were redefined for reasons
    of clarity). The corresponding loads $L_c(t)$ are available
    from model simulations for different weekdays and seasons \cite[p.\ 154]{Klobasa2007}. This data is used for estimation of the
    potential for demand side management in Germany as shown in
    Figures \ref{fig:example-klobasa-winter} and \ref{fig:example-klobasa-summer}.}
\end{table}

\begin{figure}
  \centering
  \includegraphics[width=.99\textwidth]{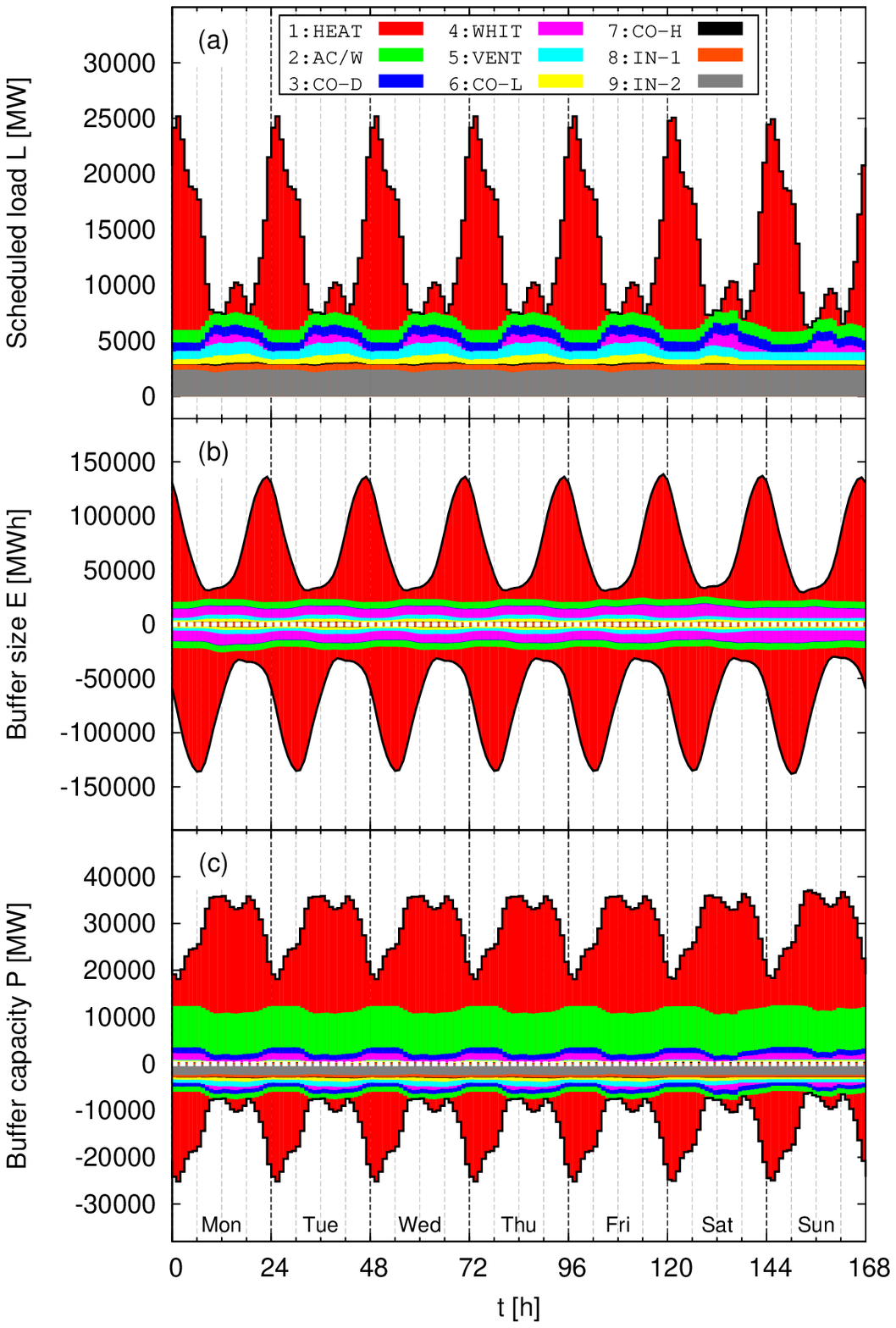}
  \caption{\label{fig:example-klobasa-winter}Exemplified application
    of the framework for characterization of the potential for demand
    side management to the accumulated loads eligible for DSM in
    Germany during a \emph{winter} week. The colors in the respective
    panes indicate contributions from the respective categories
    detailed in Table \ref{tab:klobasa-categories}.  The scheduled
    loads in panel (a) were obtained from \cite[p.\
    154]{Klobasa2007}. Panels (b) and (c) indicate the contributions
    of the respective categories to envelopes of the sizes and
    capacities of the respective buffers as defined in Equations
    (\ref{eq:e-and-p-definitions}). In winter, the potential for DSM
    is strongly dominated by electrical heating devices. Due to their
    limited shifting potential of only $8~\mathrm{h}$ the size of the
    associated storage-equivalent buffer fluctuates significantly in
    time.  Please note, that the conditions (\ref{eq:validity-r-size})
    and (\ref{eq:validity-r-capacity}) need to be taken into account
    for each individual category and that a graphical validation of
    realized loads therefore is not possible from this accumulated
    plot.  Individuals plots for the respective categories (just as
    e.g.\ presented in Figure \ref{fig:example-1-schematic}) would be
    required for this purpose instead.}
\end{figure}

\begin{figure}
  \centering
  \includegraphics[width=.99\textwidth]{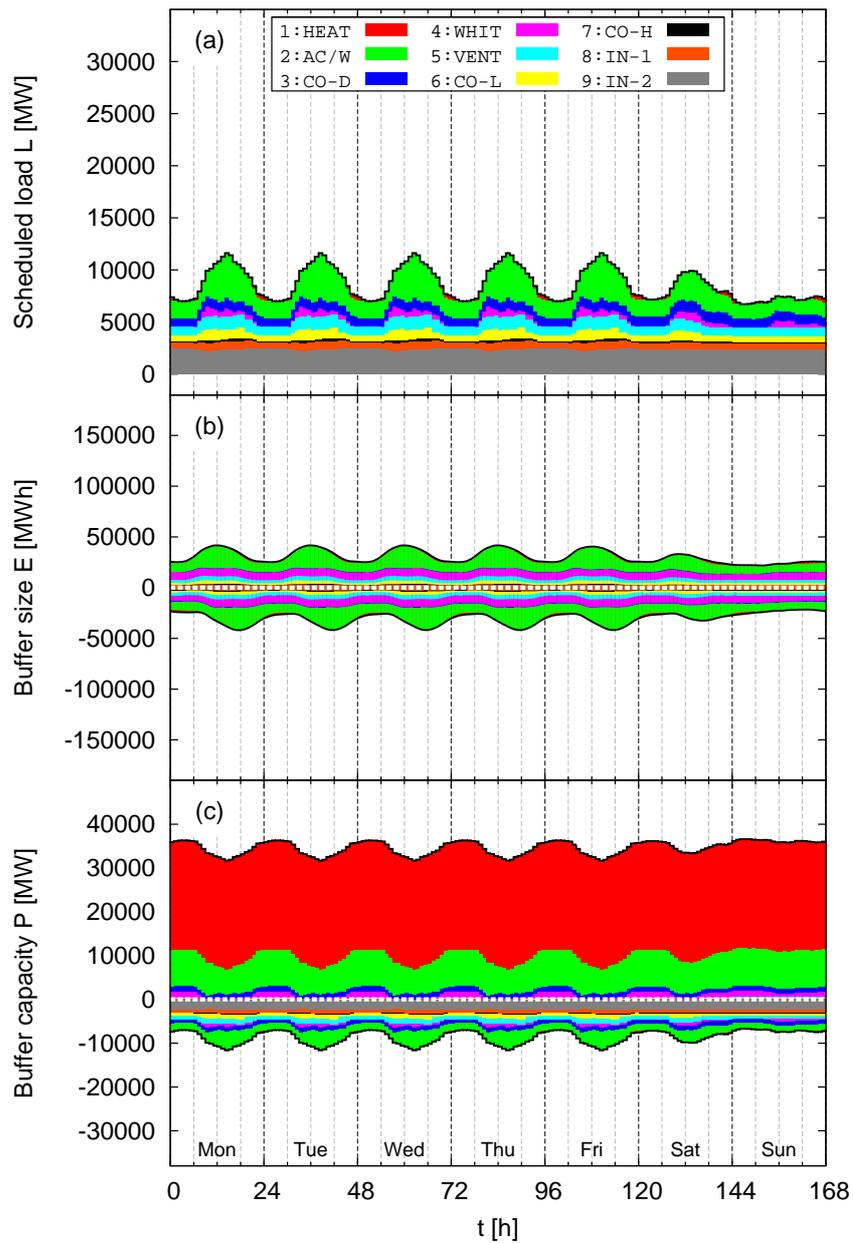}
  \caption{\label{fig:example-klobasa-summer}Exemplified application
    of the framework for characterization of the potential for demand
    side management to the accumulated loads eligible for DSM in
    Germany during a \emph{summer} week.  For details of the
    representation it is referred to the caption of Figure
    \ref{fig:example-klobasa-winter}. In summer, the overall part of
    the demand available for DSM is much lower than in winter, since
    there is no significant contribution of electrical heating.
    Instead, air conditioning and warm water (category 2) gain in
    importance.}
\end{figure}

As a second example the framework introduced in Section
\ref{sec:novel-fram-quant} is used for the conceptual characterization
of the potential for DSM in Germany.  An adequate categorization of
loads with respect to their DSM characteristics in $C=9$ categories
and corresponding scheduled loads $L_c(t)$ is obtained from
\cite{Klobasa2007}, where an in-depth analysis of the current German
electricity sector and results from model simulations are
presented. For this purpose Klobasa analyzed relevant industrial
processes and domestic consumers with respect to their potential for
the temporal shifting of loads. In combination with information on the
dissemination of the respective technologies characteristic factors
for DSM potential of different sectors of German electricity
consumption could be estimated \cite[p.\ 133]{Klobasa2007}. Klobasa
furthermore provides simulation data of the time-resolved loads of the
respective sectors, both for a summer and a winter week \cite[p.\
154]{Klobasa2007}.

Details of the categories relevant for this work are compiled in Table
\ref{tab:klobasa-categories}. Simulation results for sector-specific
load curves are available for weekdays, Saturdays and Sundays both in
winter and summer. For the scope of this analysis the data was
rearranged to cover one week (i.e.\ 5 weekdays, Saturday and Sunday)
with periodic boundary conditions\footnote{Periodic boundary
  conditions were implemented by investigating a sequence of three
  weeks in the respective seasons. Data for the innermost week was
  then used for analysis.} in winter and summer, respectively. The
maximum capacities in the respective categories were set to the
constant values listed in Table \ref{tab:klobasa-categories}.

The data and results for the analysis with the framework introduced in
Section \ref{sec:novel-fram-quant} are exhibited in Figures
\ref{fig:example-klobasa-winter} (winter week) and
\ref{fig:example-klobasa-summer} (summer week). From inspection of the
figures in particular the high potential of electric heating systems
in winter for DSM actions becomes evident.

To our knowledge this time-dependent size, which is highly relevant
for the characterization of the potential of DSM as energy
storage-equivalent technology in energy system models, was not
investigated before. As far as the capacities for charge and discharge
are concerned the peak values are in line with the estimates available
from other contributions (see e.g.\ \cite{Stadler2008a,Stotzer2011}).
Generally first and foremost categories $1$, $2$ and $4$ seem to be
appropriate for short-time management purposes, since only these
categories come with significant buffer sizes and capacities.  From
inspection of panel (c) of Figures \ref{fig:example-klobasa-winter}
and \ref{fig:example-klobasa-summer}, however, in particular for
categories 1 and 4 limitations in the provision of positive control
power (which is equivalent with a discharge of the buffer) become
evident, which originate from the intermittent operation of the
corresponding devices.

In winter, electrical heating provides an enormous potential for
DSM. With existing technologies size and capacity of the associated
storage could be increased straightforwardly by either increasing the
number of electrical heating devices or by increasing the size of
attached heat storages.  The latter would result in an increase of
$\Delta t_1$, which in particular would imply less fluctuations in the
buffer size in time. If this flexibility would already be considered
with scheduling the loads, increased time frames for management
formally could even result in increased opportunities for the
provision of positive control power. These effects would significantly
ease the applicability of DSM as storage-equivalent technology.

In summer the potential for DSM is significantly lower (see Figure
\ref{fig:example-klobasa-summer}), with the highest share in category
$2$ (air conditioning and warm water).  Both applications, however, in
principle could be attached to larger reservoirs, which significantly
would increase the size of the associated energy buffer.

The analyzed data set does not make use of the entire flexibility
of the framework, since time-dependent maximum loads $\Lambda_c$
(differing e.g.\ between summer and winter) are not provided for the
respective categories. Instead the constant values listed in Table
\ref{tab:klobasa-categories} are used. For this reason the lower
panels of Figures \ref{fig:example-klobasa-winter} and
\ref{fig:example-klobasa-summer} might overestimate the maximum
charging rates e.g.\ for electrical heating in summer. In this
particular example the effect seems to be insignificant, since the
associated energy buffer for category 1 anyway is very small during
summer and the overall effect of electrical heating therefore is not
relevant. Generally the flexibility of considering temporal (or
seasonal) fluctuations in the availability of devices, however, should
be taken into account.

We would like to draw the attention to the fact, that in Figures
\ref{fig:example-klobasa-winter} and \ref{fig:example-klobasa-summer}
buffer sizes and capacities for all categories are accumulated in
single plots, since they are relevant for a discussion of the
potential for contributions to DSM from the individual categories. A
graphical validation of realized loads as shown in Figure
\ref{fig:example-1-schematic} is, however, not feasible in this
representation, since the constraints (\ref{eq:validity-r}) separately
need to be fulfilled for each of the nine categories. For this purpose
instead individual plots for the respective categories -- just as
shown in Figure \ref{fig:example-1-schematic} -- would be required.

\section{\label{sec:conclusions-outlook}Conclusions}
In the scope of this contribution a novel, consistent framework for
the characterization of DSM as storage-equivalent technology was
introduced with the aim to ease the integration of DSM in energy
system models and unit commitment schemes.

The framework first and foremost addresses the modeling of loads,
which without significant reservation can be shifted in
time.\footnote{The integration of loads suitable for
  \emph{curtailment} is addressed in Appendix
  \ref{sec:integration-class-2}.} Traditionally DSM for these
application is implemented in terms of the optimization of a
redistribution of individual loads. However, differing requirements
for individuals loads impede a graphical illustration of the potential
for DSM, its integration into large-scale system models and the
optimization of complex commitment schemes.

With the novel framework developed in Section
\ref{sec:novel-fram-quant} an intermediate degree of complexity is
introduced instead: DSM is treated both with respect to energy and
capacity balances. From scheduled load curves and the corresponding
capacities in Equations (\ref{eq:e-and-p-definitions}) envelopes of
storage-equivalent energy buffers and their respective capacities for
charge and discharge are defined. At this level of abstraction the
operation of individual devices and their individual time frames of
management do not need to be considered. Instead it is sufficient to
consider conditions summarized in Equations (\ref{eq:validity-r}) in
conceptual energy system models and large-scale unit commitment
strategies. Technical aspects concerning the small-scale dispatch of
individual devices then can be addressed later independently from the
desired gross effects at large-scale. For reasons of clarity and
generality the Equations in Section \ref{sec:novel-fram-quant} treat
the time $t$ as a continuous variable. A time-discrete formulation is,
however, included in Appendix \ref{sec:time-discr-form}.

In Section \ref{sec:examples} the application of the framework was
discussed based on two examples. The first example, a single load
pulse, is mainly interesting for conceptual reasons. On this example
the graphical validation of Equations (\ref{eq:validity-r}) is
demonstrated. In the second example the current DSM potential in
Germany is investigated. The approach applies the framework developed
in the scope of this work to simulation data compiled and published
previously by Klobasa \cite{Klobasa2007}. From the application of this
framework the accumulated current potential for DSM in Germany can be
estimated, which is likely to increase even further with an increased
dissemination of electric vehicles not considered yet. From inspection
of Figure \ref{fig:example-klobasa-winter} in particular the transient
size of the buffer associated with electrical heating systems becomes
evident. The data used for analysis gives a first indication, that the
storage-equivalent size of the DSM buffer for electrical heating at
times can accumulate to more than $100~\mathrm{GWh}$, which is in the
order of magnitude of the demand for short-term storage in
fully-renewable energy systems.\footnote{Rasmussen et al.\ estimated
  the demand for short-term energy storage in a European energy system
  based on renewable sources to $2~\mathrm{TWh}$ \cite{Rasmussen2012}.
  Scaled down to Germany this would imply a demand of approximately
  $400~\mathrm{GWh}$.} The transient behavior is due to the rather
limited management time frame of only $\Delta t_1=8~\mathrm{h}$, which
stems from the time when DSM was used for economic operation and
commitment of huge power plants. From a technological point of view it
would be straightforward to extend this time frame, e.g.\ by
installation of larger heat storages or by investments in the energy
efficiency of facilities. From a grid perspective this might be
worthwhile if DSM is operated as alternatives to energy storages and
obtains systemic importance.

Concluding, the novel framework developed in the scope of this work is
applicable for the characterization of the potential of DSM. The
resulting buffer sizes and capacities significantly ease the
investigation of conceptional problems in energy system modeling due
to the reduced level of detail required during simulation and
optimization. Finally the storage-equivalent character of DSM actions
becomes clear with sizes and capacities directly comparable to the
storage properties of alternative technologies.

\appendix
\section{\label{sec:integration-class-2}Integration of class 2 DSM}
In the context of this contribution the focus was on class 1 DSM,
where DSM does not effect the long term energy balance. This excludes
e.g.\ curtailment of loads. The resulting framework, however,
straightforwardly can be generalized to also include class 2 DSM
actions.

In the Equations introduced in Section \ref{sec:novel-fram-quant}, the
constraint of the energy balance is reflected by Equation
(\ref{eq:validity-r-size}). Formally this constraint can be relaxed
for class 2 DSM. Equation (\ref{eq:filling-level}) then may diverge in
the course of time, reflecting the accumulated energy surplus or
deficit due to class 2 DSM actions.

\section{\label{sec:time-discr-form}Time-discrete formulation of the
  simulation equations}
In Section \ref{sec:novel-fram-quant} for reasons of clarity the
functions $L_c(t)$ and $\Lambda_c(t)$ were assumed to be available at
arbitrary, continuous $t$. In many examples of practical relevance,
however, data instead is available at discrete time steps only. Here
we re-formulate some equations from Section \ref{sec:novel-fram-quant}
with respect to data available at discrete time only.

We therefore assume that the scheduled load is available at discrete
times $t^i=t^0+i\tau$ with $i=0,\ldots,I$. Then for each category $c$
scheduled loads $L_c^i$ with $i=0,...,I-1$ need to be specified, where
$L^c_i$ is the load scheduled in the time period from $t^i$ to
$t^{i+1}$ in units of MW. Accordingly the load capacities
$\Lambda_c^i$ are defined. From this discrete data, in turn time
continuous functions can be defined as
\begin{subequations}
  \begin{eqnarray}
    \label{eq:load-disc-to-cont}
    L_c(t)&:=&\sum_{i=0}^{I-1}\begin{cases}
      L_c^i& t^{i}\le t< t^{i+1}\\
      0&\mbox{otherwise}
    \end{cases}\quad\mbox{and}\\
    \label{eq:capacity-disc-to-cont}
    \Lambda_c(t)&:=&\sum_{i=0}^{I-1}\begin{cases}
      \Lambda_c^i& t^{i}\le t< t^{i+1}\\
      0&\mbox{otherwise}
    \end{cases}\quad.
  \end{eqnarray}
\end{subequations}
In principle these functions now can be used with the Equations
developed in the cope of Section \ref{sec:novel-fram-quant}. In fact,
the integrals defined in Equations (\ref{eq:e-max-ts}) and
(\ref{eq:e-min-ts}), however, can be evaluated for the case of
stepwise constant data. Instead of these equations the envelope of the
buffer size can be estimated directly from the discrete data as
\begin{subequations}
  \begin{eqnarray}
    \label{eq:e-max-ts-disc}E^{\max}_c(t)&:=&\sum_{i=0}^{I-1}\begin{cases}
      0&t+\Delta t_c\le t^{i}\\
      \left(t+\Delta t_c-t^{i}\right) L_c^{i}&t\le t^{i}\ \mbox{and}\ t^i<t+\Delta t_c<t^{i+1}\\
      \tau L_c^{i}&t\le t^{i}\ \mbox{and}\ t^{i+1}\le t+\Delta t_c\\
      \min(t^{i+1}-t,\Delta t_c)L_c^{i}&t^{i}<t<t^{i+1}\\
      0&t^{i+1}\le t
    \end{cases}\\
    \label{eq:e-min-ts-disc}E^{\min}_c(t)&:=&\sum_{i=0}^{I-1}\begin{cases}
      0&t\le t^{i}\\
      \left(t-t^{i}\right) L_c^{i}&t-\Delta t_c\le t^{i}\ \mbox{and}\ t^i<t<t^{i+1}\\
      \tau L_c^{i}&t-\Delta t_c\le t^{i}\ \mbox{and}\ t^{i+1}\le t\\
      \min(t^{i+1}-t,\Delta t_c)L_c^{i}&t^{i}<t-\Delta t_c<t^{i+1}\\
      0&t^{i+1}\le t-\Delta t_c
    \end{cases}
  \end{eqnarray}
\end{subequations}
With these expressions replacing Equations (\ref{eq:e-max-ts}) and
(\ref{eq:e-min-ts}) the evaluation of integrals is not necessary for
the estimation of the envelopes for buffer sizes and capacities. The
set of Equations for discrete data can be extended if also the
realized loads $R_c(t)$ are discrete, which, however, in practice
generally is not the case.

\section*{Acknowledgments}
The author kindly acknowledges discussions with and instructive
comments on the manuscript by Konrad Meyer, Stefan Weitemeyer, Arjuna
Nebel, Thomas Vogt, and three anonymous reviewers. Mitavachan Hiremath
contributed to the preparation of the load profiles used in Section
\ref{sec:char-nati-potent}. Funding of the joint project RESTORE 2050
(funding code 03SF0439A) was kindly provided by the German Federal
Ministry of Education and Research through the funding initiative
Energy Storage.


\end{document}